\documentclass[aps,prc,reprint,superscriptaddress,amsmath,amssymb]{revtex4-2}

\usepackage{setspace}
\usepackage{macros}
\usepackage{graphicx}
\usepackage{bm}
\usepackage{hyperref}
\usepackage{aasmacros}
\begin{document}


\title{Strength measurement of the $E_{\alpha}^{lab}$~=~830~keV resonance in $^{22}\rm{Ne}(\alpha,n)^{25}\rm{Mg}$ reaction using a stilbene detector}



\author{Shahina}
\author{R.J.~deBoer}
\author{J.~G\"orres}
\affiliation{Department of Physics and Astronomy, University of Notre Dame, Notre Dame, Indiana 46556, USA}
\author{R.~Fang}
\affiliation{Department of Physics and Astronomy, University of Notre Dame, Notre Dame, Indiana 46556, USA}
\author{M.~Febbraro}
\affiliation{Oak Ridge National Laboratory, Oak Ridge, Tennessee 37830, USA}
\affiliation{
Air Force Institute of Technology, Wright-Patterson Air Force Base, 45433, OH, USA}
\author{R.~Kelmar}
\affiliation{Department of Physics and Astronomy, University of Notre Dame, Notre Dame, Indiana 46556, USA}
\author{M.~Matney}
\author{K.~Manukyan}
\affiliation{Department of Physics and Astronomy, University of Notre Dame, Notre Dame, Indiana 46556, USA}
\author{J.T.~Nattress}
\affiliation{Oak Ridge National Laboratory, Oak Ridge, Tennessee 37830, USA}
\author{E.~Robles}
\affiliation{Department of Physics and Astronomy, University of Notre Dame, Notre Dame, Indiana 46556, USA}
\author{T.J.~Ruland}
\author{T.T.~King}
\affiliation{Oak Ridge National Laboratory, Oak Ridge, Tennessee 37830, USA}
\author{A.~Sanchez}
\affiliation{Department of Physics and Astronomy, University of Notre Dame, Notre Dame, Indiana 46556, USA}
\author{R.S.~Sidhu}
\affiliation{School of Physics and Astronomy, The University of Edinburgh, EH9 3FD Edinburgh, United Kingdom}
\author{E.~Stech}
\author{M.~Wiescher}
\affiliation{Department of Physics and Astronomy, University of Notre Dame, Notre Dame, Indiana 46556, USA}

\date{\today}

\begin{abstract}
The interplay between the $^{22}$Ne$(\alpha,\gamma)^{26}$Mg and the competing $^{22}$Ne$(\alpha,n)^{25}$Mg reactions determines the efficiency of the latter as a neutron source at the temperatures of stellar helium burning. In both cases, the rates are dominated by the $\alpha$-cluster resonance at 830~keV. This resonance plays a particularly important role in determining the strength of the neutron flux for both the weak and main $s$-process as well as the $n$-process. Recent experimental studies based on transfer reactions suggest that the neutron and $\gamma$-ray strengths for this resonance are approximately equal. In this study, the $^{22}$Ne$(\alpha,n)^{25}$Mg resonance strength has been remeasured and found to be similar to the previous direct studies. This reinforces an 830~keV resonance strength that is approximately a factor of three larger for the $^{22}$Ne$(\alpha,n)^{25}$Mg reaction than for the $^{22}$Ne$(\alpha,\gamma)^{26}$Mg reaction. 
 
\end{abstract}

\pacs{}

\maketitle


\section{Introduction}

The $^{22}$Ne$(\alpha,n)^{25}$Mg reaction is the main neutron source for the weak $s$-process, which occurs at the end of core helium burning in red giant stars~\cite{Pignatari_2010}. It also plays an important role as the secondary neutron source for the main $s$-process during the helium flash in the hydrogen-helium intershell regions of AGB stars~\cite{Bisterzo_2015}. In addition, the reaction is considered to be a neutron source for the $n$-process, which is triggered by the expanding supernova shockfront traversing through the helium-burning shell of pre-supernova stars~\cite{Bla76}. In all these cases, the $^{22}$Ne abundance in the helium-enriched burning environment is produced by the $^{14}$N($\alpha,\gamma$)$^{18}$F($\beta^+\nu$)$^{18}$O($\alpha,\gamma$)$^{22}$Ne reaction chain, starting from the $^{14}$N ashes of the preceding CNO hydrogen burning phase~\cite{Kaeppeler_1994}, but the exact amount of neutrons released from the $^{22}$Ne$(\alpha,n)^{25}$Mg reaction remains an open question. For this reason, the reaction rate of this neutron source has been of great interest and the focus of many direct and indirect experimental, as well as theoretical, studies over the last thirty years~\cite{Ashery, Haas, Wolke, Harms, Drotleff_91, Jaeger, Hunt, Massimi, Massimi12, Giesen, Ugalde, Talwar, Jayatissa, Ota20, Ota21, Lotay, Koehler, Adsley}.

However, in order to determine the neutron production from the $^{22}$Ne$(\alpha,n)^{25}$Mg reaction it is not enough to just determine its cross section, the cross section of the radiative capture reaction $^{22}$Ne$(\alpha,\gamma)^{26}$Mg, the only other significant decay channel open at low energies, must also be determined. The role of the $^{22}$Ne$(\alpha,\gamma)^{26}$Mg reaction is further enhanced because the $^{22}$Ne$(\alpha,n)^{25}$Mg reaction has a negative $Q$-value of $Q_{(\alpha,n)}$~=~-478~keV, thus this neutron-production channel opens only under higher temperature conditions. On the other hand, the $^{22}$Ne$(\alpha,\gamma)^{26}$Mg reaction has a positive $Q$-value, $Q_{(\alpha,\gamma)}$~=~10614~keV, thus it can contribute to the gradual depletion of $^{22}$Ne at lower temperatures before the $^{22}$Ne$(\alpha,n)^{25}$Mg reaction can be significantly accessed~\cite{Kaeppeler_1994, Longland, Talwar, Ads21}. 

Further, recent studies underline the claim that the rate of both of these reactions, for all of these environments, primarily depends on the strength of the $\alpha$-cluster resonance at $E_{\alpha}^{lab}$~=~830~keV. A number of indirect studies seem to have ruled out significant contributions from lower energy resonances as summarized by \citet{Wiescher_2023}, at least for the $^{22}$Ne$(\alpha,n)^{25}$Mg reaction. The resonance strength in the radiative capture branch has recently been confirmed to be $\omega\gamma$~=~35$\pm$~4~$\mu$eV by \citet{Shahina}, in excellent agreement with previous studies~\cite{Wolke, Giesen}. Although it should be noted that there is some disagreement regarding the $E_p$~=~479~keV $^{22}$Ne$(p,\gamma)^{23}$Na resonance strength~\cite{Depalo, Williams} used for the absolute normalization of \citet{Shahina}. While there are still uncertainties with respect to the low energy resonant contributions in the radiative capture reaction, at temperatures above $T\approx$~0.25~GK - corresponding to the conditions in the helium flash driving the main $s$-process and the $n$-process in the supernova shockfront - the rate for both reactions will be dominated by the $E_{\alpha}^{lab}$~=~830~keV resonance. 

Direct measurements of the $^{22}$Ne$(\alpha,n)^{25}$Mg resonance strength are less consistent, yielding a range of values for $\omega\gamma_{(\alpha,n)}$ between 80(30) and 234(77)~$\mu$eV~\cite{Harms, Giesen, Drotleff_91, Drotleff_93,Jaeger}. The large spread in the resonance strengths and their uncertainties seem to be primarily due to systematic errors in determining the neutron detection efficiencies and background contributions in, for the most part, moderator type detectors. Recent $^{22}$Ne($^6$Li$,d$)$^{25}$Mg and $^{22}$Ne($^7$Li$,t$)$^{25}$Mg $\alpha$-transfer studies selectively populated the level at $E_x\approx$~11.320~MeV, which corresponds to the resonance in the $^{22}$Ne+$\alpha$ reactions at $E_{\alpha}^{lab}$~=~830~keV, and measured the $\gamma$-ray and neutron partial decay branching in coincidence with deuterium or tritium, respectively~\cite{Jayatissa, Ota20, Ota21}. Based on these results, a substantially smaller value for the $^{22}$Ne$(\alpha,n)^{25}$Mg resonance strength was deduced. This is in strong disagreement with the direct measurements of the resonance strength and would significantly reduce the role of the $^{22}$Ne$(\alpha,n)^{25}$Mg reaction as a stellar neutron source. 

With these motivations in mind, a new experiment was performed to independently determine the $(\alpha,n)$ resonance strength of the $E_{\alpha}^{lab}$~=~830~keV resonance in the $\Nean$ reaction, complementing the recent direct $(\alpha,\gamma)$~\cite{Shahina}. In the following section, the experimental details of this study are summarized, namely the setup and the choice of neutron detectors. In Sec.~\ref{sec::exp_proc_analysis}, the calculation of the resonance strength from the experimental observations is described and in Sec.~\ref{sec::disc} the present resonance strength is compared with previous values. A summary is given in Sec.~\ref{sec::sum}.  

\section{Experimental Details} \label{sec::exp_setup}

This section details the experimental setup, target properties, and characteristics of the detector utilized for the present study of the $E_{\alpha}^{lab}$~=~830~keV resonance in the $^{22}$Ne$(\alpha,n)^{25}$Mg reaction.

\subsection{Experimental Setup} \label{subsec::acclerator}

\begin{figure}
   \centering
    \includegraphics[width=0.5\textwidth]{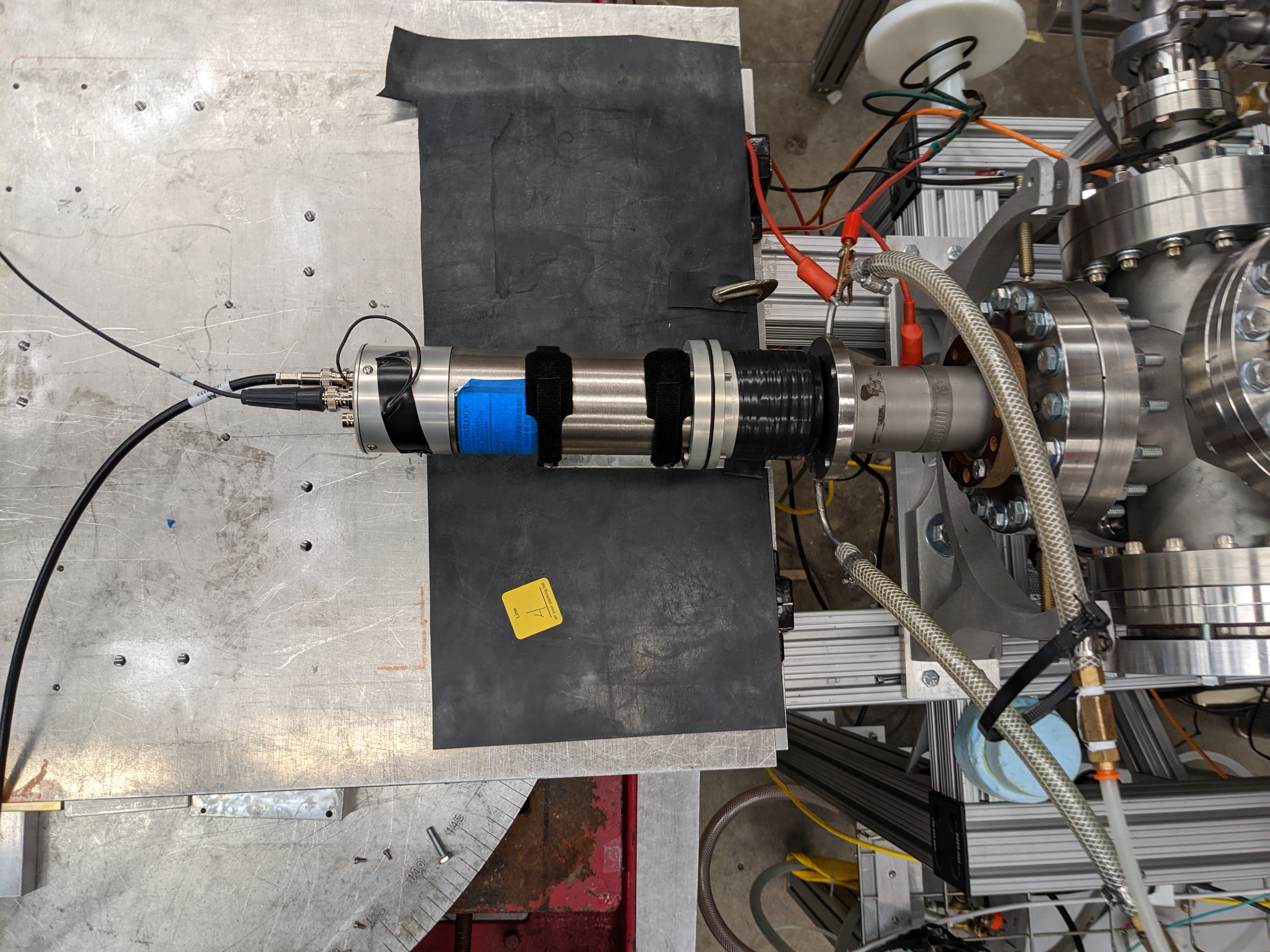}
  \caption{Experimental setup for the measurement of the $E_{\alpha}^{lab}$~=~830~keV resonance strength in the $\Nean$ reaction. The stilbene detector was placed directly behind the target holder to maximize detection efficiency.}
  \label{near_geometry_setup}
\end{figure}

Measurements were made at the University of Notre Dame Nuclear Science Laboratory~\cite{doi:10.1080/10619127.2014.882732} using the St.~ANA Pelletron accelerator. The 5~MV accelerator delivered proton and $\alpha$-particle ($^4$He$^+$) beams with an intensity of up to 65~$\mu$A on target. The energy calibration of the accelerator was determined by measuring the well-known resonances in the $^{19}\rm{F}(p,\alpha\gamma)^{16}\rm{O}$ reaction at $E_p$~=~872~keV~\cite{PhysRevC.103.055815} and the $^{27}$Al$(p,\gamma)^{28}$Si reaction at $E_p$~=~992~keV~\cite{Antilla1977}. The beam was sent through a 90$^\circ$ analyzing magnet before being transported to the experimental target station, providing a beam energy uncertainty of $\pm$1~keV at 1~MeV. The beam was defocused and wobbled over an area of approximately 1~cm~$\times$~1~cm on the target to mitigate target degradation and reduce the effects of target inhomogeneities. The target was mounted at the end of the beam line as shown in Fig.~\ref{near_geometry_setup}. A stilbene organic scintillator detector was used to measure the outgoing neutrons from the $^{22}$Ne$(\alpha,n)^{25}$Mg reaction and was placed at 0$^\circ$ relative to the beam direction, directly behind the target as shown in Fig.~\ref{near_geometry_setup}. The electrically isolated target chamber acted as a Faraday cup, which was used for the integration of the beam current. To limit carbon buildup on the target, a cold trap was mounted that consisted of a copper tube, cooled to liquid nitrogen temperature, placed inside the beam line. The copper tube extended to within a few millimeters of the target surface. The cold trap was electrically isolated and biased to $-$200~V to suppress secondary electrons. The target was water-cooled to reduce damage resulting from beam induced power deposition.

\subsection{Target Properties} \label{sec::targets}

Beam-stop backings of Ta with a thickness of 0.5~mm were used for all targets. The primary target was one with deeply-implanted $^{22}$Ne. Similarly, deeply-implanted $^{20}$Ne targets were used for background characterization. Thin $^{13}$C and natural abundance LiF targets were created by evaporation for background and stilbene detector characterization. Table~\ref{tab:targets} summarizes target and reaction properties relevant for the present measurements.  
\begin{table*}
\caption{\label{tab:targets} List of targets used during the experiment. }
\begin{ruledtabular}
\begin{tabular}{c c c c c c}
Target & Backing & Reaction & $Q$-value (keV) & $E^{lab}$ (keV) & $E_n$ (keV) ($\theta_{\text{lab}}$~=~0$^{\circ}$)\\
\hline
$^{13}\rm{C}$ & Ta & $\Can$ & 2216 & 910 & 3120\\
LiF & Ta & $\Lipn$ & -1644 & 2059 & 278\\
$^{51}$V & Ta & $^{51}$V$(p,n)^{51}$Cr & -1535 & 1835 & 285 \\
$^{20}\rm{Ne}$ & Ta & $^{20}\rm{Ne}(\alpha,n)^{23}\rm{Mg}$ & -7215 & 830 & $-$ \\
$^{22}\rm{Ne}$ & Ta & $\Nean$ & -478	& 830 & 285 \\
\end{tabular}
 \end{ruledtabular}
\end{table*}

The deeply implanted $\Ne$ targets were similar to those used by \citet{Shahina}. The $\Ne$ targets were produced by implanting $\Ne$ ions onto a 0.5~mm thick Ta backing with a dosage of 300~mC over an area of 3~$\rm{cm^{2}}$ at an energy of 200~keV. The distribution of the implanted $\Ne$ atoms in the Ta backing was measured using the well-known resonance in the $\Nepg$ reaction at $E_p^{lab}$~=~1278~keV (see \citet{Shahina}). Because of the implantation profile, a beam energy of $E_{\alpha}^{lab}$~=~910~keV was required to reach the central portion of the thick-target yield plateau for the narrow $E_{\alpha}^{lab}$~=~830~keV resonance ($\Gamma <$~3~keV~\cite{Jaeger}). The thick-target yield on the plateau was constant to $\pm$10\% over an energy range of $\pm$30~keV from this central energy. The $\Ne$ target stability was monitored throughout the experiment by measuring the yield of the strong resonance at $E_{\alpha}^{lab}$~=~1434~keV in the $\Nean$ reaction. No decrease in the yield was observed; hence the target proved stable even after accumulating a charge of nearly 3~C. 



Despite the use of a cold trap, which greatly reduced the amount of carbon build-up on the target, carbon deposition still resulted in substantial background from the $\Can$ reaction. To simulate this background source for characterization, a thin $^{13}$C target was used. The target was fabricated at the Institute for Nuclear Research (ATOMKI) in Debrecen, Hungary, using enriched (99\%) $^{13}$C powder evaporated onto the Ta backing, creating a thin layer of $\approx$10~$\mu$g/cm$^2$. As these measurements were only used to obtain the shape of the light output spectrum (see Sec.~\ref{subsec::stilbene}), the precise target thickness was not needed.

To characterize the light output spectrum as a function of mono-energetic neutron energy (see Sec.~\ref{subsec::stilbene}), the well-known $^7$Li$(p,n)^7$Be reaction was utilized. An evaporated LiF target, with a thickness of 100(7)~$\mu$g/cm$^2$, was fabricated by evaporation. The target thickness was determined by measuring the energy loss from the thick-target yield over the $E_p$~=~873~keV resonance ($\Gamma\approx$~4.3~keV~\cite{Becker1982}) in the $^{19}$F$(p,\alpha\gamma)^{16}$O reaction and stopping powers from SRIM~\cite{SRIM}. 



\subsection{Stilbene detector} \label{subsec::stilbene}

Measurement were made using an Inradoptics Scintinel cylindrical stilbene crystal 2 inches in diameter $\times$ 2 inches in length. Stilbene has a lower scintillation efficiency than standard liquid scintillators but provides excellent pulse shape discrimination (PSD), which represents the detector’s ability to discriminate between $\gamma$-ray and neutron events at lower light output (i.e., neutron energy). This was critical for this experiment because, at the $E_{\alpha}^{lab}$~=~830~keV resonance in the $^{22}$Ne$(\alpha,n)^{25}$Mg reaction, neutrons were produced with a maximum outgoing energy of $E_n\approx$~285~keV at 0$^\circ$, well below the threshold of many types of standard liquid scintillators. 

All of the scintillation 
signals were digitized using an 8-channel, 14-bit, 500-MS/s waveform digitizer (CAEN DT5730) using DPP-PSD (Digital Pulse Processing for Charge Integration and Pulse Shape Discrimination) firmware~\cite{CAENCoMPASS}. The trigger time, integrated charge, and a pulse shape parameter were recorded for each event. The integrated charge was recorded into two time windows, a long-gate integral ($Q_{long}$) and short-gate integral ($Q_{short}$), relative to the pulse trigger. The waveform record was 992~ns long. The CAEN Multi-PArameter Spectroscopy Software (CoMPASS) was used for recording~\cite{CAENCoMPASS}. 

The pulse shape discrimination (PSD) was calculated for each signal waveform using the digitizer's onboard field-programmable gate array to perform pulse-shape analysis. In this way, minimal data was sent to disk for each event, maximizing the digitizer's withstandable event rate. PSD was defined as the difference between $Q_{long}$ and $Q_{short}$ divided by $Q_{long}$, given by
\begin{equation}\label{PSP}
\mathrm{PSD} = \left(Q_{long}-Q_{short}\right)/Q_{long}.
\end{equation}
For the stilbene detector, $Q_{long}$ and $Q_{short}$ were determined as the total charge in the time windows [$t_s$, $t_s$+500~ns] and [$t_s$, $t_s$+80~ns], respectively, with a reference delay of $t_s=50$~ns after the leading edge trigger.

\begin{figure*}
   \centering
    \includegraphics[width=1\textwidth]{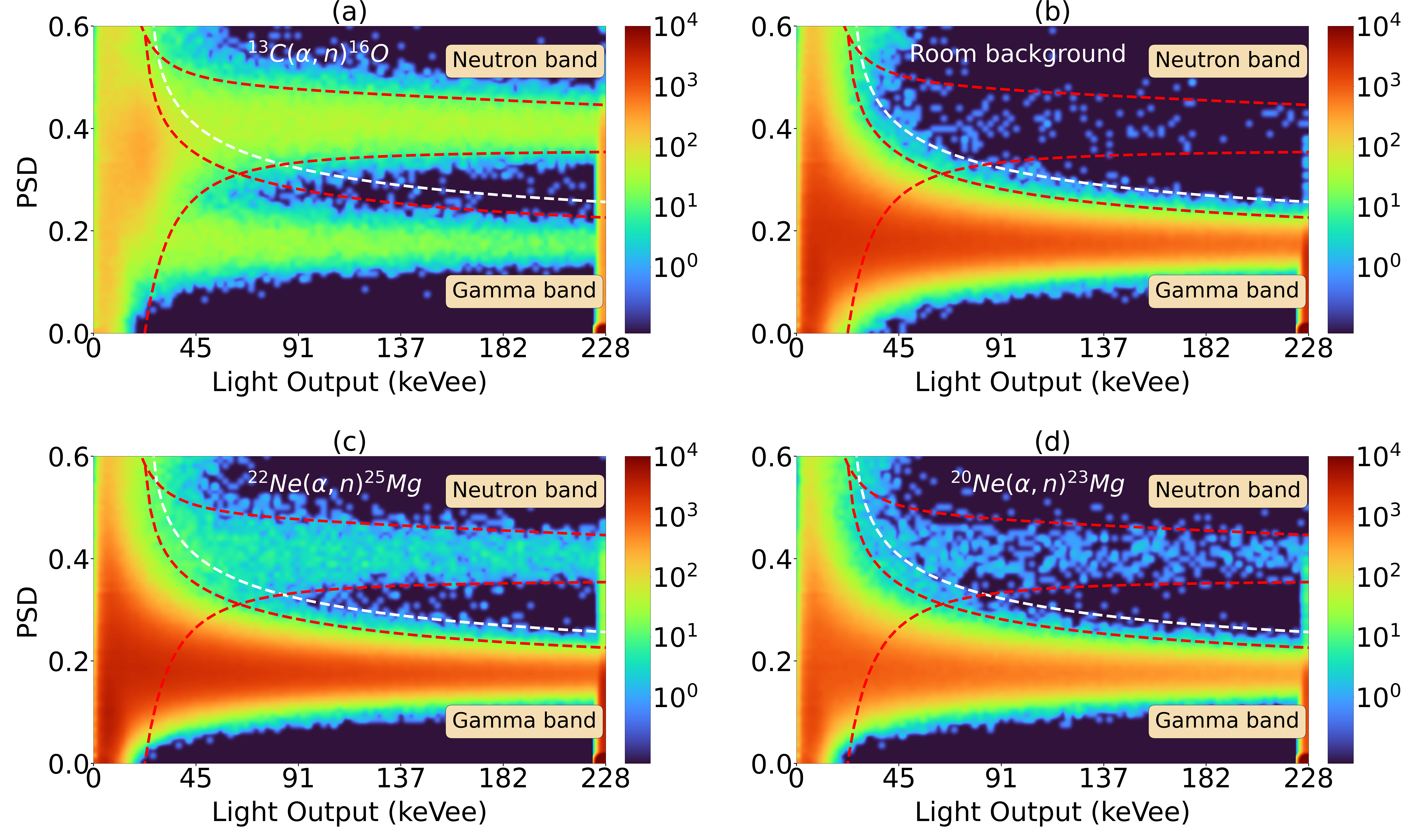}
  \caption{PSD as a function of light output (keVee) from the stilbene scintillator neutron detector for (a) an $\alpha$-particle beam on a $^{13}$C target, (b) room background, (c) an $\alpha$-particle beam on a $^{22}$Ne target and (d) an $\alpha$-particle beam on a $^{20}$Ne target used for background characterization. All targets had a 0.5~mm Ta backing. The region bounded by the three red dashed lines defines the neutron gate. The white dashed line represents the shifted gamma-ray upper bound (see Sec.~\ref{subsec::22Nean_spectrum_analysis} for details)}.
  \label{2D_hists_all}
\end{figure*}

A clear distinction between pulses generated by $\gamma$-rays and neutrons can be seen in the PSD spectra of Fig.~\ref{2D_hists_all}. The lower band corresponds to the $\gamma$-ray events, and the upper band corresponds to the neutron events. The neutron gate is bounded by the three dashed red lines, representing the $\gamma$-ray upper bound, the neutron lower bound, and the neutron upper bound. The bands were defined by fitting a function of the form $\frac{a}{\sqrt{L}} + bL + c$, in which the $a\sqrt{L}$ term represents the statistical fluctuation of photons. The $\gamma$-ray upper bound was obtained from the room background spectrum as shown in Fig.~\ref{2D_hists_all} (b), and the neutron upper and lower bands were obtained from the high statistics $\Can$ spectrum shown in Fig.~\ref{2D_hists_all} (a). The $\gamma$-ray upper bound also denotes the PSD threshold.

\begin{figure*}
   \centering
    \includegraphics[width=1.0\textwidth]{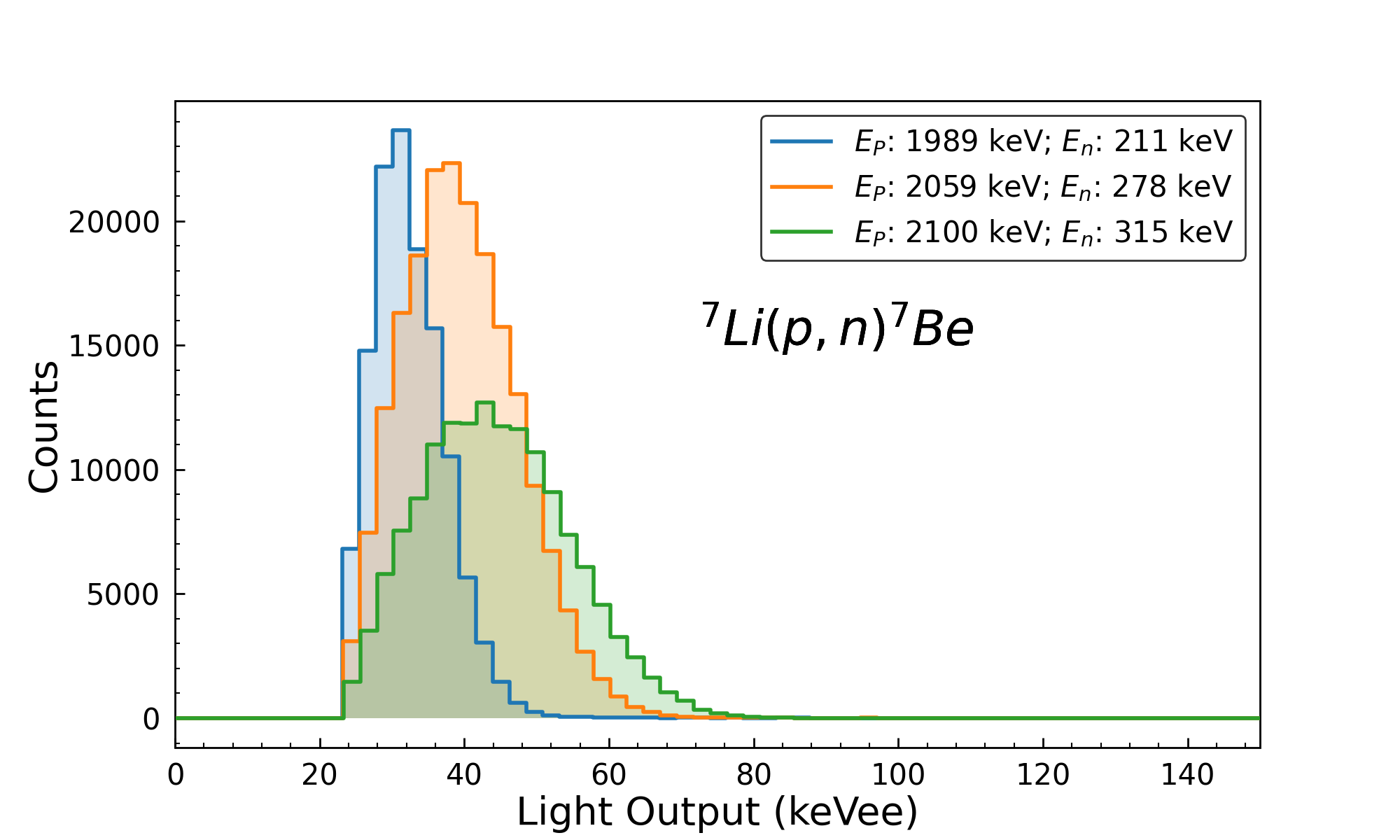}
  \caption{Example light output spectra from the stilbene detector produced by neutrons from the $^7$Li$(p,n)^7$Be reaction at $E_p$~=~1.989, 2.059, 2.100~MeV which correspond to $E_n$~=~211, 278, and 315~keV at $\theta_{lab}$~=~0$^\circ$, respectively.}
  \label{Li_spectrum}
\end{figure*}

Once the neutron bands were clearly defined, the light output spectra produced by the stilbene detector for mono-energetic neutrons over the energy range of interest was characterized using the $^7$Li$(p,n)^7$Be reaction with the target described in Sec.~\ref{sec::targets}. Three example spectra are shown in Fig.~\ref{Li_spectrum}, where protons with energies of $E_p$~=~1.989, 2.059, 2.101~MeV were used to produce neutrons of energies $E_n$~=~211, 278, and 315~keV, respectively, at $\theta_{lab}$~=~0$^\circ$. These light output spectra demonstrate how the shape evolves near the neutron energy of the $E_\alpha$~=~830~keV resonance in the $^{22}$Ne$(\alpha,n)^{25}$Mg reaction ($E_n$~=~285~keV).

\section{Analysis} \label{sec::exp_proc_analysis}

In this section, the different components required to calculate the strength of the $E_{\alpha}^{lab}$~=~830~keV resonance for the $\Nean$ reaction is discussed. First, in Sec.~\ref{subsec::det_eff}, the determination of the efficiency of the stilbene detector, using the well known $\Lipn$ reaction, is described. Then, in Sec.~\ref{subsec::background_charcterization}, the various neutron background contributions are characterized and in Sec.~\ref{subsec::22Nean_spectrum_analysis} the extraction of the neutron events from the $\Nean$ reaction is described. The spin-parity of the $E_{\alpha}^{lab}$ =~830~keV resonance, along with the angular distribution of the outgoing neutrons, are discussed in Sec.~\ref{subsec::angular_dist}. Finally, the resonance strength is calculated in Sec.~\ref{subsec::resonance_strength_calc}.

\subsection{Detector efficiency} \label{subsec::det_eff}

The efficiency $\epsilon(E_{n})$ of the stilbene detector, in the setup shown in Fig.~\ref{near_geometry_setup}, was determined by independently using the $^7$Li$(p,n)^7$Be and $^{51}$V$(p,n)^{51}$Cr reactions. This approach helps to provide a cross check on the systematic uncertainties associated with each of the two measurements. 

\subsubsection{Detector efficiency based on the $^7$Li$(p,n)^7$Be reaction study}\label{subsubsec::Lipn}

\begin{figure*}
   \centering
    \includegraphics[width=1.0\textwidth]{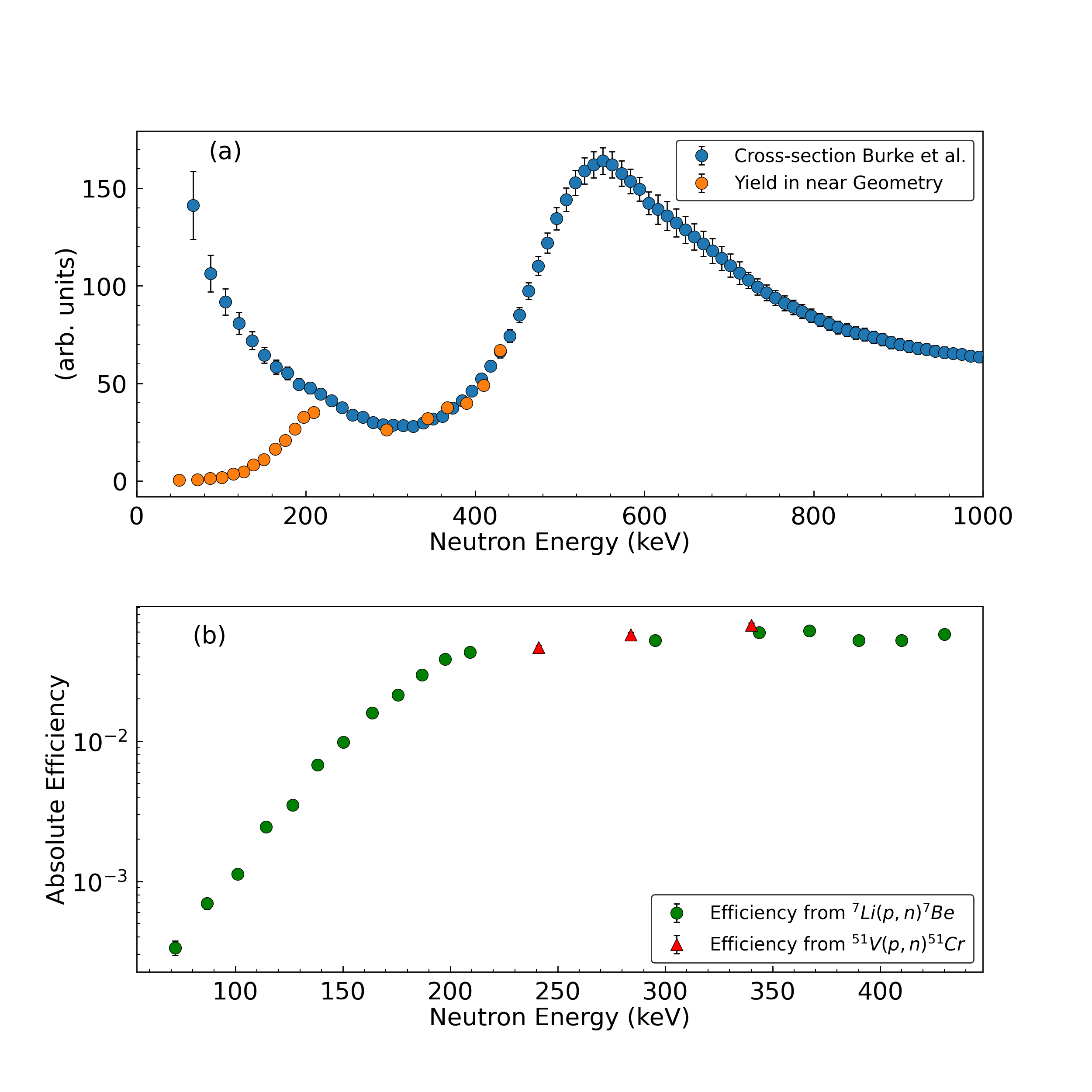}
  \caption{(a) The shape of the yield from the $\Lipn$ (orange circles) as a function of neutron energy compared with the shape of the zero degree differential cross section (blue circles) of \citet{Burke1974}. (b) The green data points represent the neutron detector efficiency obtained from the ratio of the neutron yield to the laboratory frame differential cross section of \citet{Burke1974} at $\theta_{lab}$~=~0$^\circ$ from the $\Lipn$ reaction as explained in Sec.~\ref{subsubsec::Lipn}. The red data points represent the efficiency obtained from the $^{51}$V$(p,n)^{51}$Cr reaction using the activation method. Note that the plotted data have not been corrected for the extended geometry of the detection setup (see Sec.~\ref{subsec::neutron transport}).}
  
  \label{yield_curve_near}
\end{figure*}

The neutron detection efficiency was determined using the well-known $\Lipn$ reaction~\cite{LISKIEN197557}, utilizing the target described in Sec.~\ref{sec::targets}, by taking the ratio of the 0$^\circ$ neutron yield from the neutron gate $dN(E_{P})/d\Omega_{lab}$ (see Sec.~\ref{subsec::stilbene}) to the 0$^\circ$ differential cross section from~\citet{Burke1974} $d\sigma/d\Omega_{lab}$
\begin{equation}
    \eta(E_{n}) = \frac{dN(E_{P})/d\Omega_{lab}}{N_bd\sigma/d\Omega_{lab}},
\end{equation}
where $N_b$ is the number of beam particles incident on the target. Fig.~\ref{yield_curve_near} (a) shows the 0$^\circ$ yield in the stilbene detector scaled to the differential cross section reported by \citet{Burke1974}. The shape of the yield and the differential cross section are very similar down to $E_n\approx$200~keV, where the PSD cuts, shown in Fig.~\ref{2D_hists_all}, begin to remove a substantial number of the neutron events from the overall spectrum. Fig.~\ref{Li_spectrum} illustrates how the light response spectrum changes as a function of neutron energy over the energy range of interest. The absolute efficiency is given in Fig.~\ref{yield_curve_near} (b), indicating that the PSD cuts do not significantly effect the efficiency above $E_n\approx$~200~keV. 


\subsubsection{Detector efficiency using the $^{51}$V$(p,n)^{51}$Cr reaction}\label{subsub::51Vpn}

The efficiency of the stilbene neutron detector was also determined by the activation method using the $^{51}$V$(p,n)^{51}$Cr reaction. The $^{51}\mathrm{V}$ targets were produced by evaporating a thin layer of vanadium onto a tantalum backing~\cite{1965NucIM..33..257H, 2013NIMPA.700...53F}. The reaction product $^{51}\mathrm{Cr}$ decays to the first excited state in $^{51}\mathrm{V}$ by electron capture and emits a characteristic $\gamma$-ray at 320~keV with a branching ratio of $9.9 \%$ and half-life of 27.7 days. The $^{51}\mathrm{V}$ targets were irradiated with protons with energies ranging from $E_{P}$ = 1.790~MeV to 1.887~MeV in order to produce isotropic neutrons from 241 to 340~keV. 

The efficiency of the detector can be obtained by taking the ratio of the number of neutrons detected ($n_{d}$) by the stilbene detector during the proton irradiation to the number of neutrons released from the reaction ($n_{r}$)
\begin{equation}
    \epsilon = \frac{n_{d}}{n_{r}},
\end{equation}
where $n_{r}$ is also equal to the number of $^{51}\mathrm{Cr}$ nuclei produced, which was determined by measuring the activity of the activated targets.

As the $^{51}\mathrm{V}$ targets are irradiated, the radioactive $^{51}\mathrm{Cr}$ that is formed undergoes radioactive decay. The rate of decay is given by $\lambda N$, where $\lambda$ is the decay constant and $N$ is the total number of radioactive nuclei present. The difference between the rate of production and the rate of decay gives the rate of change in $N$
\begin{equation}\label{eq:dNdt}
    \frac{dN}{dt} = P - \lambda N,
\end{equation}
where $P$ is the production rate, which remains nearly constant during the short irradiation times. 
After the irradiation for a time $t_{0}$ the $^{51}\mathrm{V}$ target will have an activity $A_{0}$ given by
\begin{equation}
    A_{0} = P(1-e^{-\lambda t_{0}}).
    \label{A_0}
\end{equation}

After the proton irradiation, the activated $^{51}\mathrm{V}$ target was transferred to a Pb-shielded $\gamma$-ray counting station. The decay of $^{51}\mathrm{Cr}$ produces a characteristic $\gamma$-ray at 320~keV, where the efficiency of the $\gamma$-ray detection setup was found to be 3.7(1)~\%. 

The activity is continuously decaying during this counting time. The number of emitted gamma-rays ($N_{\gamma}$) from the $^{51}\mathrm{Cr}$ during a counting interval between $t_{1}$ and $t_{2}$ is given by 
\begin{equation}
   N_{\gamma}=  \frac{N_{C}}{\epsilon_{Ge} I} =  \int_{t_{1}}^{t_{2}} A_{0} e^{-\lambda(t-t_{0})}dt
   = \frac{A_{0}}{\lambda} e^{\lambda t_{0}}(e^{\lambda t_{1}} - e^{-\lambda t_{2}}),
   \label{N_g}
\end{equation}
where $N_{C}$ is the number of recorded $\gamma$-rays, $\epsilon_{Ge}$ is the efficiency of the HPGe detector, and $I$ is the branching ratio. Combining Eq.~\ref{A_0} and Eq.~\ref{N_g} gives
\begin{equation}
    P = \frac{N_{C}}{I\times \epsilon_{Ge}} \frac{\lambda}{(1-e^{-\lambda t_{0}})e^{\lambda t_{0}} (e^{-\lambda t_{1}} - e^{-\lambda t_{2}})}.
\end{equation}
Multiplying the production rate with the activation time $t_{0}$ results in the number of neutrons released during the irradiation. The neutron detection efficiency can then be expressed as 
\begin{equation}
    \epsilon = \frac{n_{d}}{P \times t_{0}}.
\end{equation}
The efficiency from the $^{51}$V$(p,n)^{51}$Cr measurement is found to be consistent with that obtained from the $\Lipn$ data as shown in Fig.~\ref{yield_curve_near}. 

  

\subsubsection{Neutron transport simulation} \label{subsec::neutron transport}

The close proximity of the stilbene detector to the target means that the detector appends a rather large angular acceptance. Detected neutrons are averaged over the angular-dependent reaction kinematics as well as the angular distribution from the reaction cross-section. To account for the role these effects have on the detection efficiency, Monte Carlo-based neutron transport simulations were performed using MCNP6.2~\cite{TechReport_2017_LANL_LA-UR-17-29981_WernerArmstrongEtAl}. These simulations were benchmarked against the $\Lipn$ efficiency data. Simulated elastically scattered proton events were output using the particle tracking (PTRAC) output option. The events were converted to light using an empirical light response function and folded with a detector resolution function. Next, a PSD value was assigned to each event based on an empirical PSD function. All three functions were derived from the $\Lipn$ data. Simulated neutrons accounted for both angular-dependent reaction kinematics as well as the center-of-mass angular distribution from the reaction cross section for the $\Lipn$ and $\Nean$ reactions (see \citet{LISKIEN197557} and Sec.~\ref{subsec::angular_dist}, respectively), while that of the $^{51}$V$(p,n)^{51}$Cr reaction was assumed to be isotropic. For the reaction cross section, Legendre polynomials were fit to the angular distributions and used as inputs for generating the MCNP source terms. An efficiency ratio was determined that accounted for the differences in reaction kinematics and angular distributions between each reaction. This ratio for $\Lipn$/$\Nean$ was found to be 0.902(45), while for $^{51}$V$(p,n)^{51}$Cr/$\Nean$ it was 1.005(50). After applying these corrections, the efficiency of neutron detection from the close geometry setup for the $\Nean$ reaction at $E_\alpha$~=~830~keV ($E_n$~=~285~keV) was found to be 0.0588(34) and 0.0572(37)using the $\Lipn$ and $^{51}$V$(p,n)^{51}$Cr reactions, respectively.

\subsection{Background Characterization} \label{subsec::background_charcterization}
To investigate the neutron background from the $\Can$ reaction, a thin $^{13}\rm{C}$ target was utilized (see Sec.~\ref{sec::targets}). A measurement of the light output spectrum was made at the same $\alpha$-particle energy as the $^{22}$Ne thick-target yield measurements ($E_\alpha$~=~910 keV). Because the $\Can$ reaction has a positive $Q$-value of 2.2~MeV, this reaction produces neutrons with an energy of $E_n\approx$~3.1~MeV at $\theta_{\text{lab}}$~=~0$^\circ$. This neutron energy exceeds the corresponding maximum of the light output provided by the range of the electronics, so the maximum of the light output spectrum could not be observed. However, it was found that light output range was sufficient to constrain the contributions coming from both the $^{13}$C$(\alpha,n)^{16}$O and the $^{22}$Ne$(\alpha,n)^{25}$Mg reaction (see Sec.~\ref{subsec::stilbene}) as shown in Fig.~\ref{2D_hists_all} (a). 

To understand the beam-induced background, data were acquired with an enriched implanted $^{20}\rm{Ne}$ target at the same beam energy of $E_{\alpha}^{lab}$~=~910~keV, with a total accumulated charge of 1.267~C. Due to its large negative $Q$-value (-7215~keV, see Table~\ref{tab:targets}), this target should not produce neutrons from $^{20}$Ne$(\alpha,n)^{23}$Mg reaction, but as this target was created in the same manner as the $^{22}$Ne target, it should produce a similar background signal from neutron production on trace impurities introduced by the implantation process or already present in the backing material. The background spectrum from the $^{20}\rm{Ne}$ target has been characterized as shown by the two dimensional PSD plot in Fig.~\ref{2D_hists_all} (d). 


\subsection{$\Nean$ Spectrum Analysis} \label{subsec::22Nean_spectrum_analysis}

\begin{figure*}
   \centering
    \includegraphics[width=1.0\textwidth]{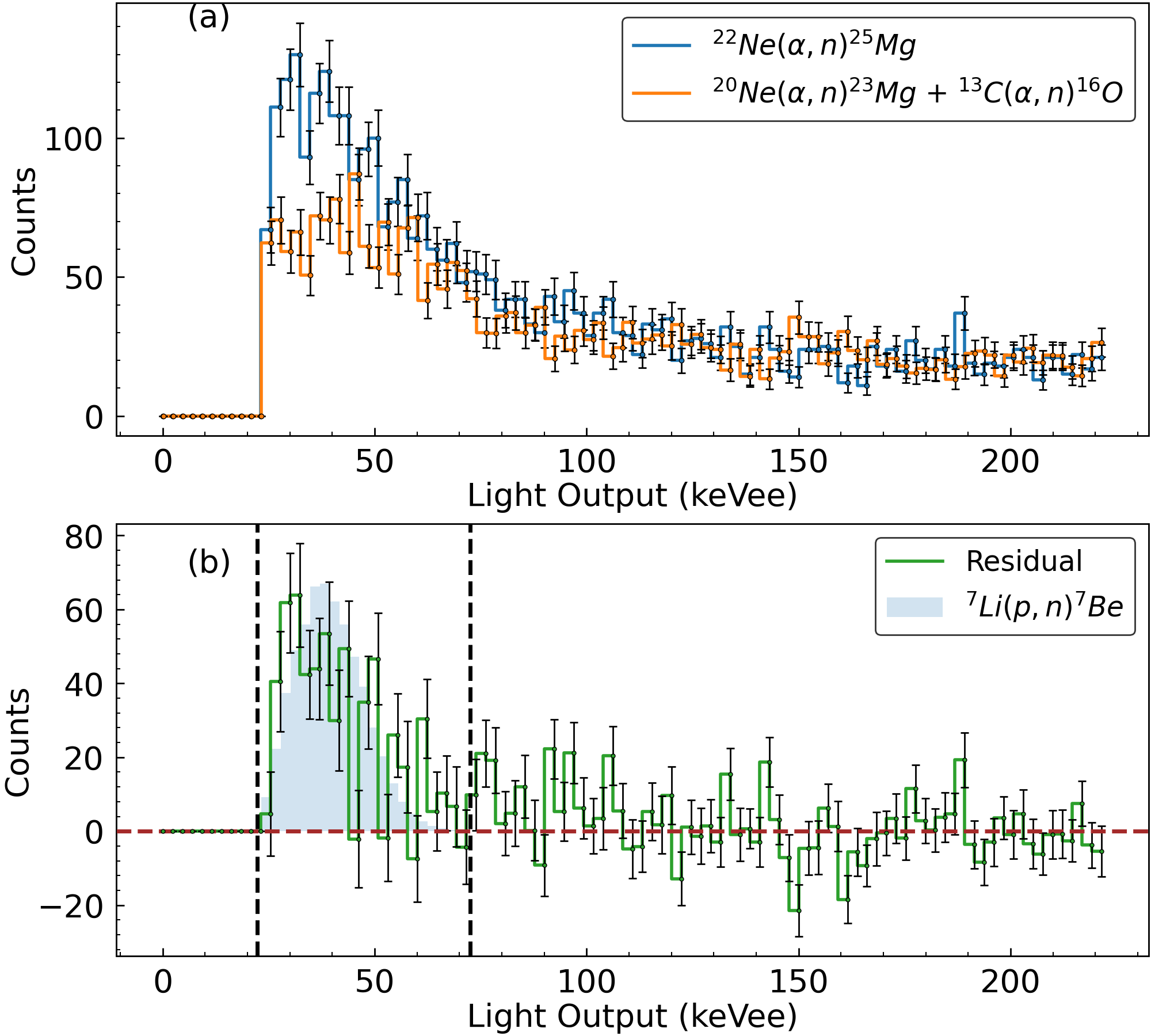}
  \caption{(a) The blue histogram shows the light output spectra from the stilbene detector produced by neutrons from the $\Nean$ reaction for the $E_{\alpha}^{lab}$~=~830~keV resonance. The orange histogram shows the sum of light output obtained from $^{20}$Ne and $^{13}$C target at the same $\alpha$-particle energy. (b) The residual plot of light output obtained after subtraction of the $\Nean$ light output (in blue) from the sum of beam-induced neutron background obtained using $^{20}$Ne target and $^{13}$C target from the $\Can$ reaction (in orange). The shaded blue histogram shows the light output from the $\Lipn$ reaction at $E_p$~=~2059~keV, which produces neutrons of a similar energy ($E_n$~=~278~keV) as that of the $\Nean$ reaction. The dashed black lines show the region of integration for the neutrons from the $\Nean$ reaction. }
  \label{1D_hist}
\end{figure*}

Data were acquired at the top of the thick-target plateau, i.e., at $E_{\alpha}^{lab}$~=~910~keV (see \citet{Shahina} and Sec.~\ref{sec::targets}), with a total accumulated charge of $Q$~=~2.997~C on the $\Ne$ target. The two dimensional PSD plot from the $\Nean$ reaction is shown in Fig.~\ref{2D_hists_all} (c). The neutron band, bounded by the three dashed red lines, defined the neutron gate used to extract the neutron events as described in Sec.~\ref{subsec::stilbene}. The 1D projection of the extracted neutron band is shown in Fig.~\ref{1D_hist} (a), where the blue histogram represents the light response spectrum for the measurement with the implanted $\Ne$ target. However, the light response spectrum contains a sizable background component, as indicated by the significant number of counts at higher light output. As shown in Fig.~\ref{Li_spectrum}, neutrons with an energy of $\approx$285~keV are expected to produce a maximum light output of $\approx$73~keVee (see Sec.~\ref{subsec::stilbene}).

To determine the number of counts corresponding to neutron events from the $\Nean$ reaction, two distinct background subtraction methods were employed. In the first approach, the background spectra obtained from $^{20}$Ne and $^{13}$C targets were used to model the higher light output background so that the background over the low light output region of the $\Nean$ spectrum could be subtracted as shown in Fig.~\ref{1D_hist} (a). The beam-induced background was determined using the $(\alpha,n)$-inactive $^{20}$Ne implanted target. The $^{20}$Ne target spectrum was normalized to the amount of charge accumulated on the $^{22}$Ne target. However, the yield observed in the high light output portion of the $^{20}$Ne light response spectrum was found to be substantially lower than that observed in the $^{22}$Ne target spectrum. This has been interpreted as resulting from increased deposition of $^{13}$C on the $^{22}$Ne target in comparison to the $^{20}$Ne target. This discrepancy is attributed to the significantly greater charge exposure on the $^{22}$Ne target ($Q$ = 2.997 C) in contrast to the $^{20}$Ne target ($Q$ = 1.267 C). Hence, the $\Can$ target spectrum was included as an additional background source, scaled using a $\chi^{2}$ fit to the high light output portion of the spectrum ( i.e., from 73-223 keVee) and added to the $^{20}$Ne target spectrum. The residual plot indicates a surplus of low energy neutrons coming from the $\Nean$ reaction while the higher light output region is on average consistent with zero as shown Fig.~\ref{1D_hist} (b). The expected shape of the light response spectrum using neutrons of a similar energy ($E_n\approx$~278~keV), obtained with high statistics from the $^7$Li$(p,n)^7$Be calibration runs, is shown for comparison. The net number of counts observed from neutrons associated with the population of the $E_{\alpha}^{lab}$~=~830~keV resonance on the thick-target yield plateau at $\theta_{lab}$~=~0$^\circ$ was found to be $dN_{max}^\infty/d\Omega_{lab}$~=~552, with a statistical uncertainty of 10\%.

\begin{figure*}
   \centering
    \includegraphics[width=1.0\textwidth]{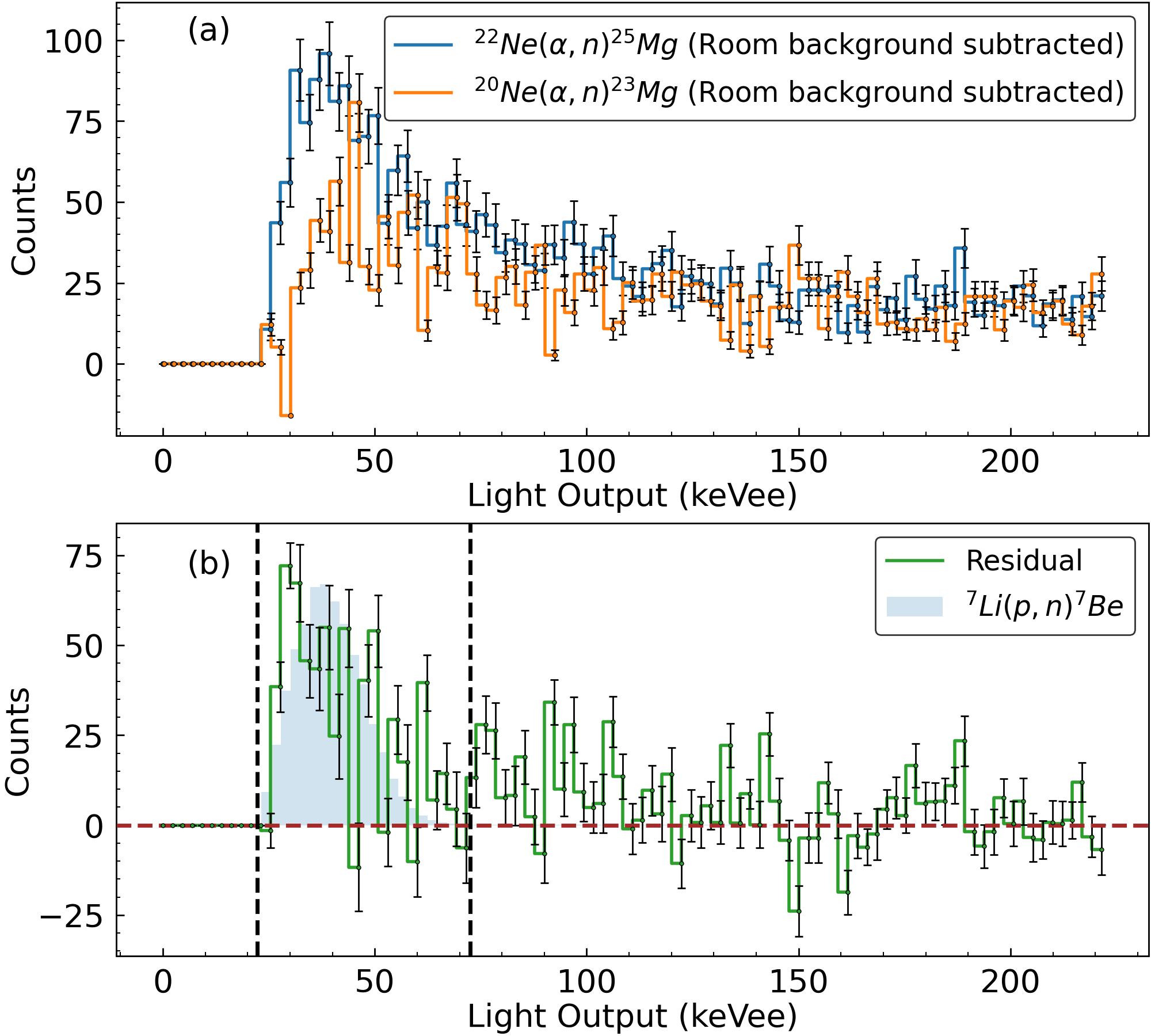}
  \caption{As Fig.~\ref{1D_hist}, except the background subtraction is done using only the room background spectrum as described in the text.}
  \label{1D_hist_background_subtracted}
\end{figure*}

To validate our results, an alternative background subtraction technique was employed. In this method, the room background was subtracted from the $^{22}$Ne and $^{20}$Ne target spectra, as depicted in Fig.~\ref{1D_hist_background_subtracted} (a). The $^{20}$Ne light response spectrum was then scaled using a $\chi^{2}$ fit to the high light output portion. The residual plot (Fig.\ref{1D_hist_background_subtracted} (b)) again revealed a surplus of low-energy neutrons from the $\Nean$ reaction, while the higher light output region remained consistent with zero. The net count of neutrons associated with the $E_{\alpha}^{lab}$~=~830~keV resonance was determined to be $\sim$~576, with a statistical uncertainty of 7.7\%. The agreement between both methodologies instills confidence in identifying the $\Can$ reaction as the primary source of background in the high light output portion. The total number of counts was then taken as the weighted average of the above two values, resulting in 567(35). Based on this and other tests of the background subtraction a conservative systematic uncertainty of 15\% has been estimated for the overall background characterization.

For the low energy neutrons under investigation, another source of background could come from $\gamma$-ray leakage into neutron band as shown in Fig.~\ref{2D_hists_all}. In order to test for incomplete $\gamma$-ray / neutron discrimination, a more restrictive gate, with the $\gamma$-ray upper bound shifted up (as shown in Fig.~\ref{2D_hists_all} by the white dashed line) was also analyzed. This more restrictive gating comes with a decrease in statistics but should be consistent with the more wider gate if $\gamma$-ray leakage is small. Indeed, when the more restrictive gate was utilized, the number of counts in the neutron band reduced to 393, a decrease of nearly 30\%. However, the efficiency given the more restrictive band was also found to decrease by 22\%. Thus the resulting yield remained consistent within the estimated uncertainties.


Neutron background from the $^{11}$B$(\alpha,n)^{14}$N reaction has also been reported previously in the low energy $^{22}$Ne$(\alpha,n)^{25}$Mg study of \citet{Harms}, but at the lower energy of $E_\alpha\approx$~600~keV. If a significant yield of neutrons was present from this reaction, the light output spectrum would show a non-linear shape at higher light output, as the end point peak of the neutron spectrum ($E_n\approx$950~keV) would fall within the range of the acquisition system. Further, it is estimated that the contributing yield from the $^{11}$B$(\alpha,n)^{14}$N reaction should be quite weak at the energy of measurement, since in the $^{22}$Ne$(\alpha,\gamma)^{26}$Mg measurements of \citet{Shahina}, which used nearly identical targets, significant yield from the $^{11}$B$(\alpha,\gamma)^{15}$N reaction was only observed at the lower energy runs at $E_\alpha$~=~650~keV, closer to the strong resonance in the $^{11}$B$(\alpha,\gamma)^{15}$N and $^{11}$B$(\alpha,n)^{15}$N reactions at $E_\alpha^{lab}$~=~606.0(5)~keV~\cite{Wang1991}.

\subsection{$\Nean$, $E_\alpha$~=~830~keV Resonance Angular Distribution } \label{subsec::angular_dist}
There is some debate~\cite{Wiescher_2023, Hunt, Ota20, Chen} as to the spin-parity of the $E_{\alpha}^{lab}$~=~830~keV resonance, but suggested values are limited to $J^\pi$~=~0$^+$, 1$^-$, or 2$^+$. Simplifying the present analysis, the angular distribution $W(\theta_{c.m.})$ for $J^\pi$~=~0$^+$ and 2$^+$ is isotropic if the lowest allowed angular momentum channels are assumed to dominate, a good approximation at the low energies of both the entrance and exit channels. If $J^\pi$~=~1$^-$, as asserted by \citet{Wiescher_2023}, then $W(\theta_{c.m.}) = 1 + 0.2 P_2(\cos{\theta_{c.m.}})$~\cite{Biedenharn}, and $W(\theta_{c.m.}=0)$~=~1.2. However, the close geometry and large solid angle coverage of the detector setup will tend to reduce the anisotropy, meaning that a value of $W(\theta_{c.m.}=0)$~=~1.2 is an upper limit. 
A value of $W(\theta_{c.m.}=0)$~=~1.1(1) is thus conservatively adopted, under the assumption that the level corresponding to the $E_{\alpha}^{lab}$~=~830~keV resonance is $J^\pi$~=~1$^-$.


\subsection{$\Nean$, $E_\alpha$~=~830~keV Resonance Strength} \label{subsec::resonance_strength_calc}
The $(\alpha,n)$ strength ($\omega\gamma_{(\alpha,n)}$) of the $E_{\alpha}^{lab}$~=~830~keV resonance is obtained by
\begin{equation}
    \omega \gamma_{(\alpha,n)} = \frac{2 (dN_{max}^\infty/d\Omega_{lab}) (d\Omega_{lab}/d\Omega_{c.m.})\epsilon_{\mathrm{eff}}}{\lambda^{2}N_b\eta(E_n) W(\theta_{c.m.})},
    \label{eq:1}
\end{equation}
$\lambda$ is the deBroglie wavelength of the incident $\alpha$-particle in the center-of-mass frame, and $(d\Omega_{lab}/d\Omega_{c.m.})$ denotes the solid angle ratio in the laboratory to the center-of-mass reference frame. The effective stopping power of $\alpha$-particles in the implanted $^{22}$Ne target ($\epsilon_{\mathrm{eff}}$) in the center-of-mass frame is given by
\begin{equation}
    \epsilon_{\mathrm{eff}} =\frac{M_{Ne}}{M_{\alpha} +M_{Ne}} \left[\epsilon_{Ne} + \left(\frac{N_{Ta}}{N_{Ne}}\right)\epsilon_{Ta}\right],
    \label{stoi}
\end{equation}
where $N_{Ta}/N_{Ne}$ = 2.7 $\pm$ 0.3 is the initial target stoichiometry~\cite{Shahina}, $M_{\alpha}$ and $M_{Ne}$ are the masses of the projectile and target nuclei (in amu), while $\epsilon_{Ne}$ and $\epsilon_{Ta}$ denote the stopping power of $\alpha$-particles in Ne and Ta, respectively. The stopping powers are from SRIM~\cite{SRIM}. The total number of $\Ne$ nuclei from implantation on Ta is (6.21 $\pm$ 0.37) $\times$ $10^{17}$ $\mathrm{atoms/cm^{2}}$ (see \citet{Shahina}). The value of the resonance strength is then $\omega \gamma_{(\alpha,n)}$~=~100~$\pm$~22~$\mu$eV, where the uncertainty contributions are summarized in Table~\ref{tab:incert_neut}. 

\begin{table}
\caption{\label{tab:incert_neut}Sources of uncertainty for the measurement of the $E_{\alpha}^{lab}$~=~830~keV resonance strength in the $\Nean$ reaction.}
\begin{ruledtabular}
\begin{tabular}{lc}
Source & \% contribution \\
\hline
Statistics &  6\\
Background Subtraction & 15 \\ 

Current Integration & 3 \\ 

Neutron Detection Efficiency & 3 \\ 
Kinematic Effects (MCNP correction) & 5 \\
Angular distribution & 10 \\
Target Thickness & 6 \\ 
Target Stability & 5 \\
\hline
Total & 22 \\
\end{tabular}
\end{ruledtabular}
\end{table}

\section{Discussion} \label{sec::disc}

\begin{figure*}
   \centering
    \includegraphics[width=1.0\textwidth]{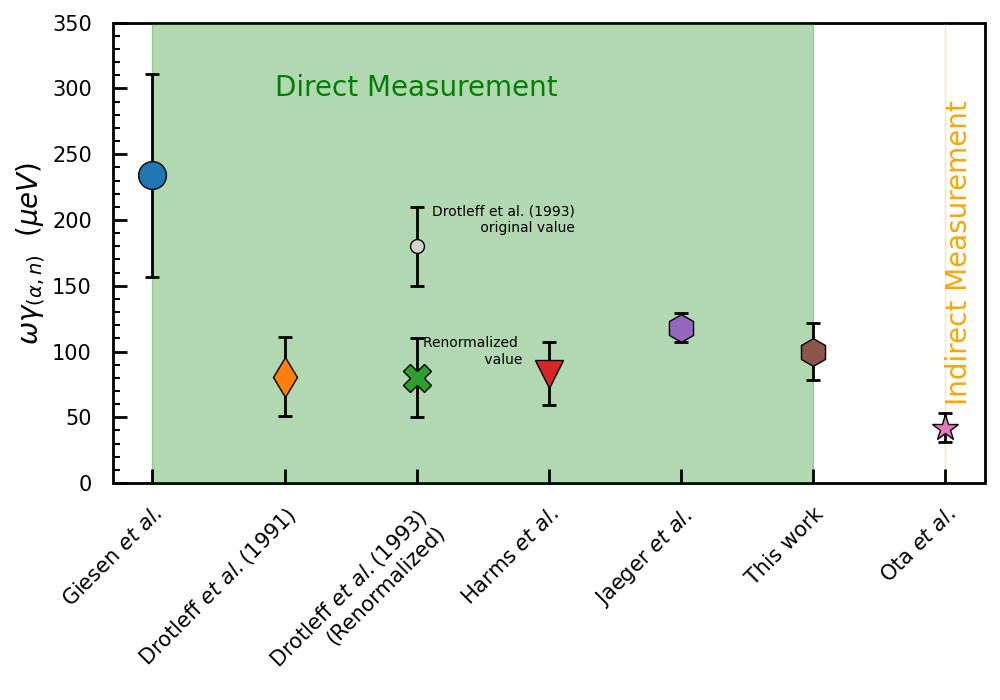}
     \caption{Comparison of $\omega\gamma_{(\alpha,n)}$ from Refs.~\cite{Giesen, Drotleff_91, Drotleff_93,Harms,Jaeger, Ota20} for the $E_{\alpha}^{lab}$~=~830~keV resonance in $\Nean$ reaction. The gray data point corresponds to the value from \citet{Drotleff_93}, where \cite{Drotleff_93} determined the strength of the $E_{\alpha}^{lab}$~=~830~keV resonance relative to the $E_{R}^{lab}$~=~1580~keV resonance. \citet{Drotleff-thesis} gives a strength of  2.9~$\pm$~0.3~eV for the 1580~keV resonance, resulting in a strength of 180~$\pm$~30~$\mathrm{\mu eV}$. However, two independent measurements by \citet{Wolke,Harms} give a strength of the 1580~keV resonance which is a factor of 2 lower. Scaling to the values of \cite{Wolke,Harms} results in a strength of 81~$\pm$~30~$\mathrm{\mu eV}$ for the \citet{Drotleff_93}, which is shown as the green data point in the plot. 
     \label{an_strength}}
\end{figure*}

Neutron resonance strengths for the $E_{\alpha}^{lab}$~=~830~keV resonance are compared in Table~\ref{results} and Fig.~\ref{an_strength} for both previous measurements~\cite{Giesen,Drotleff_91,Drotleff_93,Harms,Jaeger,Ota20} and that of the present study. All seven of these experiments were done independently with different set-up and detector arrangements over the last twenty five years. This may account for some of the fluctuations in the final results, but the large difference between some of the reported strengths warrants some more detailed discussion.


\begin{table}
\caption{Comparison of the resonance strengths from previous literature with the present work for the $E_{\alpha}^{lab}$~=~830~keV resonance in $\Nean$ reaction. \label{results} }
\begin{ruledtabular}
\begin{tabular}{c c}
Ref. & $\omega\gamma_{(\alpha,n)}$~($\mu$eV) \\
\hline
\citet{Giesen} & 234 $\pm$ 77 \\
\citet{Drotleff_91} & 80 $\pm$ 30\\
\citet{Drotleff_93}\footnotemark[1] & 81 $\pm$ 30\\
\citet{Harms} & 83 $\pm$ 24\\
\citet{Jaeger} & 118 $\pm$ 11 \\
\citet{Ota20}\footnotemark[2] & 42 $\pm$ 11 \\
This work & 100 $\pm$ 22 \\


\end{tabular}
 \end{ruledtabular}
 \footnotetext[1]{See Fig.~\ref{an_strength} text for details.}
 \footnotetext[2]{Value inferred from indirect studies.}
\end{table}

To determine the strength of the $E_{\alpha}^{lab}$~=~830~keV resonance, \citet{Giesen} used implanted $\Ne$ targets surrounded by a 4$\pi$ moderator type detector consisting of 31 $^{3}$He proportional counters embedded in polyethylene to detect the outgoing neutrons at thermal energies. The observed neutron yield showed large uncertainties, significantly amplified by beam induced background from the $^{13}$C$(\alpha,n)^{16}$O reaction caused by carbon deposition on the target surface. The neutron signal for this resonance was less than 20\% of the background yield~\cite{Giesen-thesis}. After subtracting the beam induced background estimated from the background fit, the maximum of the remaining thick-target yield curve resulted in a strength of $\omega \gamma_{(\alpha,n)}$~=~234~$\pm$~77~$\mathrm{\mu eV}$. This value is clearly considerably higher than suggested by subsequent studies, and may be the consequence of variations or inhomogeneities in the $^{13}$C buildup, or unaccounted target deterioration that led to an erroneous background subtraction. 

The reaction study by \citet{Harms} aimed primarily towards measuring the angular distribution of higher energy resonances using a windowless gas target system. For the study of the strength of the $E_{\alpha}^{lab}$~=~830~keV resonance, a combination of five $^{3}$He proportional counters and a NE213 liquid scintillator embedded in polyethylene was used~\cite{Harms-thesis}. The same windowless gas target (filled with 99.8\% enriched $\Ne$ gas) - with only minor modifications - was used by \citet{Drotleff_91}, \citet{Drotleff_93}, and \citet{Jaeger}. For neutron detection, these measurements used 4$\pi$ neutron detectors consisting of $^{3}$He proportional counters in different numbers and configurations surrounded by a polyethylene matrix. Only the result by \citet{Drotleff_93} disagrees with the other studies being nearly a factor of two higher. This appears to be a normalization error since \citet{Drotleff_93} determined the strength of the $E_{\alpha}^{lab}$~=~830~keV resonance relative to the $E_{\alpha}^{lab}$~=~1580~keV resonance. \citet{Drotleff-thesis} gives a strength of 2.9~$\pm$~0.3~eV for the 1580~keV resonance, resulting in a strength of 180~$\pm$~30~$\mathrm{\mu eV}$. In contrast, two independent measurements by \citet{Wolke} and \citet{Harms} gave a strength for the 1580~keV resonance that is about a factor of two lower. Scaling to the values of \citet{Wolke} and \citet{Harms} results in a strength of 81~$\pm$~30~$\mathrm{\mu eV}$. This renormalized value is in good agreement with the one provided by \citet{Harms} and \citet{Drotleff_91}; it also agrees within the uncertainties with the results quoted by \citet{Jaeger}. This suggests that this inconsistency with the earlier results is primarily due to a normalization error.

A recent indirect measurement by \citet{Ota20} has determined the branching ratio $\Gamma_{n}/\Gamma_{\gamma}$ or equivalently ($\omega\gamma_{(\alpha,n)}/\omega\gamma_{(\alpha,\gamma)}$) to be 1.14(26) for the $E_{\alpha}^{lab}$~=~830~keV resonance, via the $^{22}$Ne$(^{6}$Li$,d-\gamma)^{26}$Mg $\alpha$-transfer reaction in inverse kinematics. This resulted in an indirect determination of $\omega \gamma_{(\alpha,n)}$~=~42~$\pm$~11~$\mathrm{\mu eV}$, a value much smaller than any of the past direct measurements. The lack of resolution in the $^{22}$Ne$(^6$Li$,d-\gamma)$ and $^{22}$Ne$(^6$Li$,d-n)$ coincidence peak analysis shown in their figure 2 should be a challenge for a reliable analysis and even a 25\% uncertainty for their given resonance strengths seems to be optimistic. 

In this work, the low PSD threshold of the stilbene detector provided a means of detecting low energy neutrons where the endpoint of the light output spectra also gives a sensitivity to the energy of the neutron events, a capability not present in previous measurements. The strength from the present measurements is found to be $\omega\gamma_{(\alpha,n)}$~=~100~$\pm$ 22 $\mathrm{\mu eV}$. This value is in good agreement with the strength of \citet{Harms}, \citet{Drotleff_91} and \citet{Jaeger} and after corrected normalization also with \citet{Drotleff_93}. On the one hand, it is considerably larger than the strength obtained from indirect measurement~\cite{Ota20}, but also considerably smaller than the measurements of \citet{Giesen}, presumably due to faulty background subtraction.

\section{Summary} \label{sec::sum}

As one of the main neutron sources for the $s$- and the $n$-process, the cross-section of the $^{22}$Ne$(\alpha,n)^{25}$Mg reaction is a key quantity for modeling the associated nucleosynthesis environments. Recent indirect studies have suggested a significant reduction in the neutron strength of the $E_{\alpha}^{lab}$~=~830~keV resonance, which dominates the reaction rate at these temperatures. 
Such discrepancies have prompted a new measurement of $\omega\gamma_{(\alpha,n)}$ for the $E_{\alpha}^{lab}$~=~830~keV resonance. To test uncertainties due to the similar detection techniques used by previous measurements, neutron detection was accomplished using a stilbene detector, where more information on the neutron energy was retained. The resulting strength was found to be similar to that reported by \citet{Harms}, \citet{Drotleff_91} and \citet{Jaeger} and also, after renormalization, with \citet{Drotleff_93}. Combining this neutron strength with the $\gamma$-ray strength of the recent measurement of \citet{Shahina} gives $\omega\gamma_{(\alpha,n)}/\omega\gamma_{(\alpha,\gamma)}$~=~2.85(71), significantly larger than the value of 1.14(26) suggested by \citet{Ota20}.

 
\begin{acknowledgments}
We are grateful to Gy\"orgy Gy\"urky for providing the $^{13}$C target used in this research. This research was supported by the National Science Foundation through Grant No.~PHY-2011890 (University of Notre Dame Nuclear Science Laboratory), Grant No.~PHY-1430152 (the Joint Institute for Nuclear Astrophysics - Center for the Evolution of the Elements) and Grant No. OISE-1927130 (IReNA). RSS acknowledges funding from STFC (Grant No. ST/P004008/1) and the Royal Society International Exchanges Grant (IES\textbackslash R2\textbackslash232267).
\end{acknowledgments}

\bibliography{22Ne.bib}

\end{document}